\def\year{2020}\relax
\documentclass[letterpaper]{article}
\usepackage{aaai20}
\usepackage{times}
\usepackage{helvet}
\usepackage{courier}
\usepackage{amsmath, amsthm, amssymb}
\usepackage{multirow}

\usepackage{tabularx}

\usepackage{pgfplots}
\usepackage{enumitem}
\usetikzlibrary{fit,positioning}
\usepackage{subfigure}
\usepackage{booktabs}
\def\url#1{\expandafter\string\csname #1\endcsname}

\pgfplotsset{compat=1.14}

\title{A Human-AI Loop Approach for Joint Keyword Discovery and Expectation Estimation in Micropost Event Detection}

\author{Akansha Bhardwaj\textsuperscript{\rm 1},  \Large \textbf{Jie Yang\textsuperscript{\rm 2}\thanks{Work performed before joining Amazon.}}, \Large \textbf{Philippe~Cudr\'e-Mauroux\textsuperscript{\rm 1}}\\ 
\textsuperscript{\rm 1}University of Fribourg, Switzerland, \textsuperscript{\rm 2}Amazon, USA\\ 
\{akansha.bhardwaj, pcm\}@unifr.ch, jiy@amazon.com\\ 
}
\begin{document}
\maketitle
\begin{abstract}
\begin{quote}

Microblogging platforms such as Twitter are increasingly being used in 
event detection. 
Existing approaches mainly use machine learning models and rely on event-related keywords to collect the data for model training. These approaches make strong assumptions on the distribution of the relevant microposts containing the keyword -- referred to as the expectation of the distribution -- and use it as a posterior regularization parameter during model training. Such approaches are, however, limited as they fail to reliably estimate the informativeness of a keyword and its expectation for model training. This paper introduces a Human-AI loop approach to jointly discover informative keywords for model training while estimating their expectation. Our approach iteratively leverages the crowd to estimate both keyword-specific expectation and the disagreement between the crowd and the model in order to discover new keywords that are most beneficial for model training. These keywords and their expectation not only improve the resulting performance but also make the model training process more transparent. We empirically demonstrate the merits of our approach, both in terms of accuracy and interpretability, on multiple real-world datasets and show that our approach improves the state of the art by 24.3\%.

\end{quote}
\end{abstract}

\section{Introduction}

Event detection on microblogging platforms such as Twitter aims to detect events preemptively. A main task in event detection is detecting events of \textit{predetermined types}~\cite{atefeh2015survey}, such as concerts or controversial events based on microposts matching specific event descriptions. This task has extensive applications ranging from cyber security~\cite{ritter2015weakly,chambers2018detecting} to political elections~\cite{konovalov2017learning} or public health~\cite{akbari2016tweets,lee2017adverse}. Due to the high ambiguity and inconsistency of the terms used in microposts, event detection is generally performed though statistical machine learning models, which require a labeled dataset for model training. Data labeling is, however, a long, laborious, and usually costly process. 
For the case of micropost classification, though positive labels can be collected (e.g., using specific hashtags, or event-related date-time information), there is no straightforward way to generate negative labels useful for model training. To tackle this lack of negative labels and the significant manual efforts in data labeling,~\citeauthor{ritter2015weakly} (\citeyear{ritter2015weakly,konovalov2017learning}) introduced a weak supervision based learning approach, which uses only positively labeled data, accompanied by unlabeled examples by filtering microposts that contain a certain keyword indicative of the event type under consideration~(e.g., \lq hack' for cyber security). 

Another key technique in this context is expectation regularization~\cite{mann2007simple,druck2008learning,ritter2015weakly}. Here, the estimated proportion of relevant microposts in an unlabeled dataset containing a keyword is given as a \emph{keyword-specific expectation}. This expectation is used in the regularization term of the model's objective function to constrain the posterior distribution of the model predictions. By doing so, the model is trained with an \emph{expectation} on its prediction for microposts that contain the keyword. Such a method, however, suffers from two key problems:

\begin{enumerate}
    \item Due to the unpredictability of event occurrences and the constantly changing dynamics of users' posting frequency~\cite{myers2014bursty}, estimating the \emph{expectation} associated with a keyword is a challenging task, even for domain experts;
    \item The performance of the event detection model is constrained by the informativeness of the keyword used for model training. As of now, we lack a principled method for discovering new keywords and improve the model performance.
\end{enumerate}

To address the above issues, we advocate a \emph{human-AI loop} approach for discovering informative keywords and estimating their expectations reliably. Our approach iteratively leverages 1) crowd workers for estimating keyword-specific expectations, and 2) the disagreement between the model and the crowd for discovering new informative keywords. More specifically, at each iteration after we obtain a keyword-specific expectation from the crowd, we train the model using expectation regularization and select those keyword-related microposts for which the model's prediction disagrees the most with the crowd's expectation; such microposts are then presented to the crowd to identify new keywords that best explain the disagreement. By doing so, our approach identifies new keywords which convey more relevant information with respect to existing ones, thus effectively boosting model performance. By exploiting the disagreement between the model and the crowd, our approach can make efficient use of the crowd, which is of critical importance in a human-in-the-loop context~\cite{yan2011active,yang2018leveraging}. An additional advantage of our approach is that by obtaining new keywords that improve model performance over time, we are able to gain insight into how the model learns for specific event detection tasks. Such an advantage is particularly useful for event detection using complex models, e.g., deep neural networks, which are intrinsically hard to understand~\cite{ribeiro2016should,doshi2017towards}. 

An additional challenge in involving crowd workers is that their contributions are not fully reliable~\cite{vaughan2017making}. In the crowdsourcing literature, this problem is usually tackled with probabilistic latent variable models \cite{dawid1979maximum,whitehill2009whose,zheng2017truth}, which are used to perform truth inference by aggregating a redundant set of crowd contributions. Our human-AI loop approach improves the inference of keyword expectation by aggregating contributions not only from the crowd but also from the model. This, however, comes with its own challenge as the model's predictions are further dependent on the results of expectation inference, which is used for model training. To address this problem, we introduce a unified probabilistic model that seamlessly integrates expectation inference and model training, thereby allowing the former to benefit from the latter while resolving the inter-dependency between the two.

To the best of our knowledge, we are the first to propose a human-AI loop approach that iteratively improves machine learning models for event detection. In summary, our work makes the following key contributions:
\begin{itemize}
    \item A novel human-AI loop approach for micropost event detection 
    that jointly discovers informative keywords and estimates their expectation; 
    \item A unified probabilistic model that infers keyword expectation and simultaneously performs model training;
    \item An extensive empirical evaluation of our approach on multiple real-world datasets demonstrating that our approach significantly improves the state of the art by an average of 24.3\% AUC.
\end{itemize}

The rest of this paper is organized as follows. First, we present our human-AI loop approach in Section~\ref{sec:model}. Subsequently, we introduce our proposed probabilistic model in Section~\ref{UPmodel}. The experimental setup and results are presented in Section~\ref{sec:results}. Finally, we briefly cover related work in Section~\ref{sec:related} before concluding our work in Section~\ref{sec:conclusion}.

\section{The Human-AI Loop Approach}
\label{sec:model}
\begin{figure*}[t!]
\centering
\includegraphics[width=0.65\textwidth]{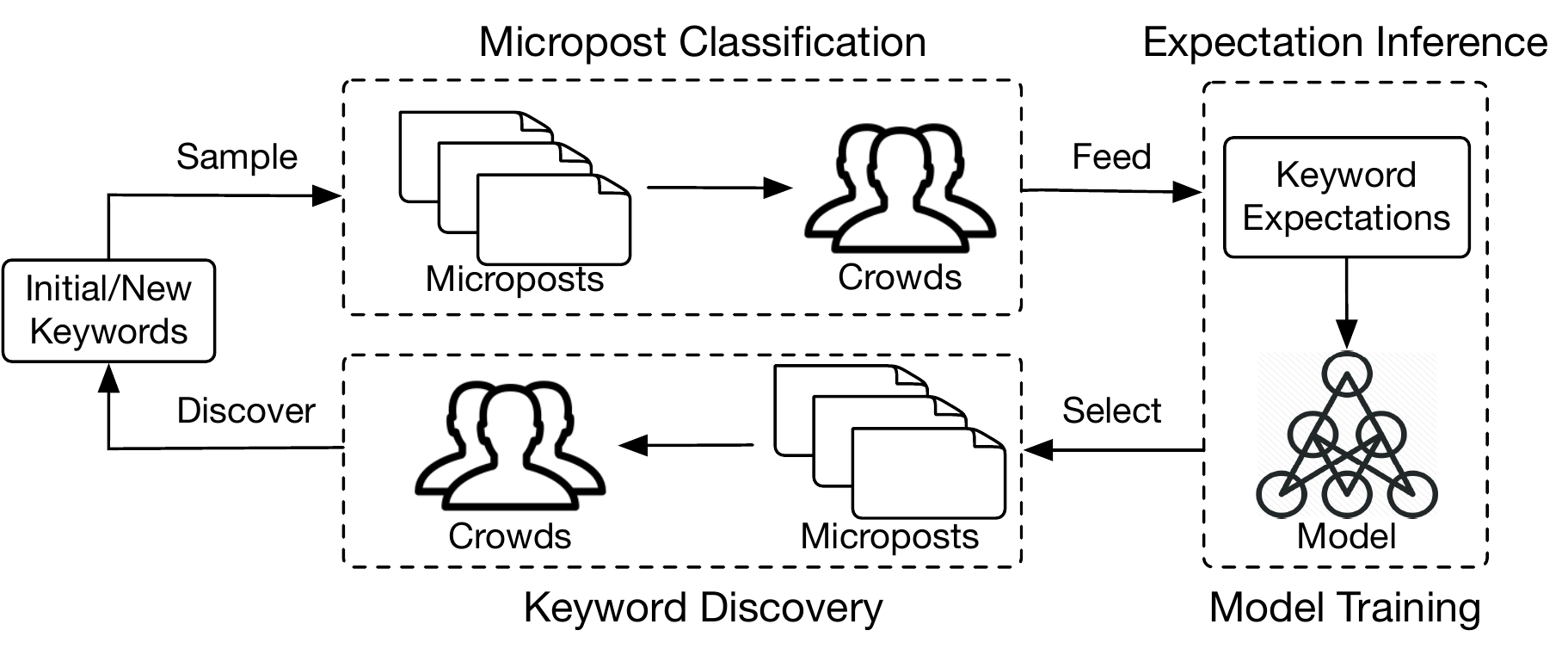}
\caption{An overview of our proposed human-AI loop approach. Starting from a (set of) new keyword(s), our approach is based on the following processes: 1) \emph{Micropost Classification}, which samples a subset of the unlabeled microposts containing the keyword and asks crowd workers to label these microposts; 2) \emph{Expectation Inference \& Model Training}, which generates a keyword-specific expectation and a micropost classification model for event detection; 3) \emph{Keyword Discovery}, which applies the trained model and calculates the disagreement between model prediction and the keyword-specific expectation for discovering new keywords, again by leveraging crowdsourcing.} 
\label{fig:system}
\end{figure*}

Given a set of labeled and unlabeled microposts, our goal is to extract informative keywords and estimate their expectations in order to train a machine learning model. To achieve this goal, our proposed human-AI loop approach comprises two crowdsourcing tasks, i.e., micropost classification followed by keyword discovery, and a unified probabilistic model for expectation inference and model training. Figure~\ref{fig:system} presents an overview of our approach. Next, we describe our approach from a process-centric perspective.

Following previous studies~\cite{ritter2015weakly,chang2016expectation,chambers2018detecting}, we collect a set of unlabeled microposts $\mathcal{U}$ from a microblogging platform and post-filter,  using an initial (set of) \emph{keyword(s)}, those microposts that are potentially relevant to an event category. Then, we collect a set of event-related microposts (i.e., positively labeled microposts) $\mathcal{L}$, post-filtering with a list of seed events. $\mathcal{U}$ and $\mathcal{L}$ are used together to train a discriminative model (e.g., a deep neural network) for classifying the relevance of microposts to an event. We denote the target model as $p_\theta(y|x)$, where $\theta$ is the model parameter to be learned and $y$ is the label of an arbitrary micropost, represented by a bag-of-words vector $x$. Our approach iterates several times $t=\{1, 2, \ldots\}$ until the performance of the target model converges. Each iteration starts from the initial keyword(s) or the new keyword(s) discovered in the previous iteration. Given such a keyword, denoted by $w^{(t)}$, the iteration starts by sampling microposts containing the keyword from $\mathcal{U}$, followed by dynamically creating micropost classification tasks and publishing them on a crowdsourcing platform.

\smallskip
\noindent\textbf{Micropost Classification.} The micropost classification task requires crowd workers to label the selected microposts into two classes: event-related and non event-related. In particular, workers are given instructions and examples to differentiate \emph{event-instance} related microposts and~\emph{general event-category} related microposts. Consider, for example, the following microposts in the context of \emph{Cyber attack} events, both containing the keyword \lq hack':
\begin{quote}
    \emph{Credit firm Equifax says 143m Americans' social security numbers exposed in hack}
\end{quote}
     This micropost describes an instance of a cyber attack event that the target model should identify. This is, therefore, an \emph{event-instance} related micropost and should be considered as a positive example. Contrast this with the following example:
     
\begin{quote}
    \emph{Companies need to step their cyber security up}
\end{quote}
This micropost, though related to cyber security in general, does not mention an instance of a cyber attack event, and is of no interest to us for event detection. This is an example of a general ~\emph{event-category} related micropost and should be considered as a negative example.

In this task, each selected micropost is labeled by multiple crowd workers. The annotations are passed to our probabilistic model for expectation inference and model training.

\smallskip
\noindent\textbf{Expectation Inference \& Model Training.} Our probabilistic model takes crowd-contributed labels and the model trained in the previous iteration as input. As output, it generates a keyword-specific expectation, denoted as $e^{(t)}$, and an improved version of the micropost classification model, denoted as $p_{\theta^{(t)}}(y|x)$. The details of our probabilistic model are given in Section~\ref{UPmodel}.
 
\smallskip
\noindent\textbf{Keyword Discovery.} The keyword discovery task aims at discovering a new keyword~(or a set of keywords) that is most informative for model training with respect to existing keywords. To this end, we first apply the current model $p_{\theta^{(t)}}(y|x)$ on the unlabeled microposts $\mathcal{U}$. For those that contain the keyword $w^{(t)}$, we calculate the disagreement between the model predictions and the keyword-specific \emph{expectation} $e^{(t)}$:
\begin{equation}
    Disagreement(x_i) = |p_{\theta^{(t)}}(y_i|x_i) - e^{(t)}|,
    \label{equ:disagree}
\end{equation}
and select the ones with the highest disagreement for keyword discovery. These selected microposts are supposed to contain information that can \emph{explain} the disagreement between the model prediction and keyword-specific expectation, and can thus provide information that is most different from the existing set of keywords for model training.

For instance, our study shows that the expectation for the keyword \lq hack' is 0.20, which means only 20\% of the initial set of microposts retrieved with the keyword are event-related. A micropost selected with the highest disagreement (Eq.~\ref{equ:disagree}), whose likelihood of being event-related as predicted by the model is $99.9\%$, is shown as an example below:

\begin{quote}
    \emph{RT @xxx: Hong Kong securities brokers hit by cyber attacks, may face more: regulator \#cyber \#security \#hacking https://t.co/rC1s9CB} 
\end{quote}

\noindent This micropost contains keywords that can better indicate the relevance to a cyber security event than the initial keyword \lq hack', e.g., \lq securities', \lq hit', and \lq attack'. 

Note that when the keyword-specific expectation $e^{(t)}$ in Equation~\ref{equ:disagree} is high, the selected microposts will be the ones that contain keywords indicating the irrelevance of the microposts to an event category. Such keywords are also useful for model training as they help improve the model's ability to identify irrelevant microposts. 

To identify new keywords in the selected microposts, we again leverage crowdsourcing, as humans are typically better than machines at providing specific explanations~\cite{mcdonnell2016relevant,chang2016crowd}. In the crowdsourcing task, workers are first asked to find those microposts where the model predictions are deemed correct. Then, from those microposts, workers are asked to find the keyword that best indicates the class of the microposts as predicted by the model. The keyword most frequently identified by the workers is then used as the initial keyword for the following iteration. In case multiple keywords are selected, e.g., the top-$N$ frequent ones, workers will be asked to perform $N$ micropost classification tasks for each keyword in the next iteration, and the model training will be performed on multiple keyword-specific expectations.
 
\section{Unified Probabilistic Model}
\label{UPmodel}

This section introduces our probabilistic model that infers keyword expectation and trains the target model simultaneously. We start by formalizing the problem and introducing our model, before describing the model learning method.

\smallskip
\noindent \textbf{Problem Formalization.} We consider the problem at iteration $t$ where the corresponding keyword is $w^{(t)}$. In the current iteration, let $\mathcal{U}^{(t)} \subset \mathcal{U}$ denote the set of all microposts containing the keyword and $\mathcal{M}^{(t)}= \{x_{m}\}_{m=1}^M\subset \mathcal{U}^{(t)}$ be the randomly selected subset of $M$ microposts labeled by $N$ crowd workers $\mathcal{C} = \{c_n\}_{n=1}^N$. The annotations form a matrix $\mathbf{A}\in \mathbb{R}^{M\times N}$ where $\mathbf{A}_{mn}$ is the label for the micropost $x_m$ contributed by crowd worker $c_n$. Our goal is to infer the keyword-specific expectation $e^{(t)}$ and train the target model by learning the model parameter $\theta^{(t)}$. An additional parameter of our probabilistic model is the reliability of crowd workers, which is essential when involving crowdsourcing. Following Dawid and Skene  \cite{dawid1979maximum,zheng2017truth}, we represent the annotation reliability of worker $c_n$ by a latent confusion matrix $\boldsymbol{\pi}^{(n)}$, where the $rs$-th element $\boldsymbol{\pi}_{rs}^{(n)}$ denotes the probability of $c_n$ labeling a micropost as class $r$ given the true class $s$. 

\subsection{Expectation as Model Posterior}
First, we introduce an expectation regularization technique for the weakly supervised learning of the target model $p_{\theta^{(t)}}(y|x)$. In this setting, the objective function of the target model is composed of two parts, corresponding to the labeled microposts $\mathcal{L}$ and the unlabeled ones $\mathcal{U}$.
 
The former part aims at maximizing the likelihood of the labeled microposts:
\begin{equation} 
    \mathcal{J}_1 = \sum_{i=1}^L \log p_\theta(y_i | x_i) + \log p_\sigma(\theta),
\end{equation}
where we assume that $\theta$ is generated from a prior distribution (e.g., Laplacian or Gaussian) parameterized by $\sigma$. 

To leverage unlabeled data for model training, we  make use of the expectations of existing keywords, i.e., \{($w^{(1)}$, $e^{(1)}$), \ldots, ($w^{(t-1)}$, $e^{(t-1)}$), ($w^{(t)}$, $e^{(t)}$)\} (Note that $e^{(t)}$ is inferred), as a regularization term to constrain model training. To do so, we first give the model's expectation for each keyword $w^{(k)}$ ($1\leq k\leq t$) as follows:
\begin{equation}
   \mathbb{E}_{x\sim \mathcal{U}^{(k)}}(y) =  \frac{1}{|\mathcal{U}^{(k)}|} \sum_{x_i\in \mathcal{U}^{(k)}} p_\theta(y_i|x_i),
\end{equation}
which denotes the empirical expectation of the model’s posterior predictions on the unlabeled microposts $\mathcal{U}^{(k)}$ containing keyword $w^{(k)}$. Expectation regularization can then be formulated as the regularization of the distance between the Bernoulli distribution parameterized by the model's expectation and the expectation of the existing keyword:
\begin{equation}
    \mathcal{J}_2 = - \lambda  \sum_{k=1}^t D_{KL}[Ber(e^{(k)})\|Ber(\mathbb{E}_{x\sim \mathcal{U}^{(k)}}(y))],
\end{equation}
where $D_{KL}[\cdot\|\cdot]$ denotes the KL-divergence between the Bernoulli distributions $Ber(e^{(k)})$ and $Ber(\mathbb{E}_{x\sim \mathcal{U}^{(k)}}(y))$, and $\lambda $ controls the strength of expectation regularization.

\subsection{Expectation as Class Prior}
To learn the keyword-specific expectation $e^{(t)}$ and the crowd worker reliability $\boldsymbol{\pi}^{(n)}$ ($1\leq n\leq N$), we model the likelihood of the crowd-contributed labels $\mathbf{A}$ as a function of these parameters. In this context, we view the expectation as the class prior, thus performing expectation inference as the learning of the class prior. By doing so, we connect expectation inference with model training. 

Specifically, we model the likelihood of an arbitrary crowd-contributed label $\mathbf{A}_{mn}$ as a mixture of multinomials where the prior is the keyword-specific expectation $e^{(t)}$:
\begin{equation}
    p(\mathbf{A}_{mn}) = \sum_s^K e_s^{(t)} \boldsymbol{\pi}_{rs}^{(n)}, 
\end{equation}
where $e_s^{(t)}$ is the probability of the ground truth label being $s$ given the keyword-specific expectation as the class prior; $K$ is the set of possible ground truth labels (binary in our context); and $r=\mathbf{A}_{mn}$ is the crowd-contributed label. Then, for an individual micropost $x_m$, the likelihood of crowd-contributed labels $\mathbf{A}_{m:}$ is given by:
\begin{equation}
    p(\mathbf{A}_{m:}) = \sum_s^K e_s^{(t)} \prod_{n=1}^N \boldsymbol{\pi}_{rs}^{(n)}.
\end{equation}
Therefore, the objective function for maximizing the likelihood of the entire annotation matrix $\mathbf{A}$ can be described as:
\begin{equation}
    \mathcal{J}_3  = \sum_{m=1}^M \log p(\mathbf{A}_{m:}).
\end{equation}

\subsection{Unified Probabilistic Model}
\label{upl}
\begin{figure}[t!]
\centering
\begin{tikzpicture}
\tikzstyle{main}=[circle, minimum size = 5.5mm, thick, draw =black!80, node distance = 7mm]
\tikzstyle{connect}=[-latex, thick]
\tikzstyle{box}=[rectangle, draw=black!100]
  \node[main, fill = white!100] (p) [label=above:$\sigma$] { };
  \node[main] (t) [below=of p,label=below:$\mathcal{\theta}$] { };
  \node[main, fill = black!10] (x) [right=of p,label=above:$x_i$] {};
  \node[main] (y) [below=of x,label=below:$y_i$] { };
  \node[main] (e) [right=of x,label=above:$e^{(t)}$] { };
  \node[main] (ym) [right=of e,label=above:$y_m$] { };
  \node[main, draw=none] (invisible) [below=of ym] { };
  \node[main, fill = black!10] (A) [right=of invisible,label=below:$\mathbf{A}_{mn}$] { };
  \node[main, draw=none] (invisible2) [right=of A] { };
  \node[main] (c) [above=of invisible2,label=above:$\boldsymbol{\pi}^{(n)}$] { };
  \path (p) edge [connect] (t)
        (t) edge [connect] (y)
		(x) edge [connect] (y)
		(e) edge [connect] (y)
		(e) edge [connect] (ym)
		(ym) edge [connect] (A)
		(c) edge [connect] (A);
  \node[rectangle, inner sep=6.4mm,draw=black!100, fit= (x) (y)] {};
  \node[rectangle, inner sep=5.6mm, draw=black!100, fit = (ym) (A) , xshift=0mm, yshift=0.6mm] {};
  \node[rectangle, inner sep=5.6mm, draw=black!100, fit = (A) (c), xshift=0mm, yshift=-0.8mm] {};
\end{tikzpicture}
\caption{Our proposed probabilistic model contains the target model (on the left) and the generative model for crowd-contributed labels (on the right), connected by keyword-specific expectation.}
\label{fig:model}
\end{figure}
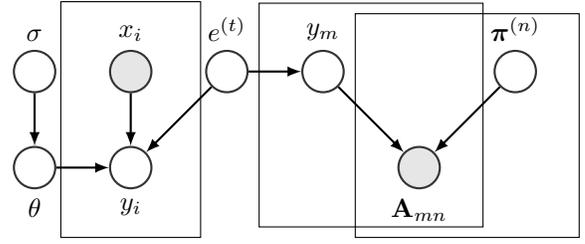

Integrating model training with expectation inference, the overall objective function of our proposed model is given by:
\begin{equation}
    \mathcal{J} = \mathcal{J}_1 + \mathcal{J}_2 + \mathcal{J}_3.
\end{equation}
Figure~\ref{fig:model} depicts a graphical representation of our model, which combines the target model for training (on the left) with the generative model for crowd-contributed labels (on the right) through a keyword-specific expectation.

\smallskip
\noindent\textbf{Model Learning.} Due to the unknown ground truth labels of crowd-annotated microposts ($y_m$ in Figure~\ref{fig:model}), we resort to expectation maximization for model learning. The learning algorithm iteratively takes two steps: the E-step and the M-step. The E-step infers the ground truth labels given the current model parameters. The M-step updates the model parameters, including the crowd reliability parameters $\boldsymbol{\pi}^{(n)}$ ($1\leq n\leq N$), the keyword-specific expectation $e^{(t)}$, and the parameter of the target model $\theta^{(t)}$. The E-step and the crowd parameter update in the M-step are similar to the Dawid-Skene model~\cite{dawid1979maximum}. The keyword expectation is inferred by taking into account both the crowd-contributed labels and the model prediction: 
\begin{equation}
    e^{(t)} \propto  \frac{1}{M}\sum_{m=1}^M y_m \times \frac{1}{|\mathcal{U}^{(t)}|} \sum_{x_i\in \mathcal{U}^{(t)}} p_\theta(y_i|x_i).
\end{equation}
The parameter of the target model is updated by gradient descent. For example, when the target model to be trained is a deep neural network, we use back-propagation with gradient descent to update the weight matrices.


\section{Experiments and Results}
\label{sec:results}
This section presents our experimental setup and results for evaluating our approach. We aim at answering the following questions:

\begin{itemize}[noitemsep,leftmargin=*]
    \item \textbf{Q1}: How effectively does our proposed human-AI loop approach enhance the state-of-the-art machine learning models for event detection?
    \item \textbf{Q2}: How well does our keyword discovery method work compare to existing keyword expansion methods?
    \item \textbf{Q3}: How effective is our approach using crowdsourcing at obtaining new keywords compared with an approach labelling microposts for model training under the same cost?
    \item \textbf{Q4}: How much benefit does our unified probabilistic model bring compared to methods that do not take crowd reliability into account? 
\end{itemize}


\subsection{Experimental Setup}

\textbf{Datasets.} We perform our experiments with two predetermined event categories: cyber security~(\emph{CyberAttack}) and death of politicians~(\emph{PoliticianDeath}). These event categories are chosen as they are representative of important event types that are of interest to many governments and companies. The need to create our own dataset was motivated by the lack of public datasets for event detection on microposts. The few available datasets do not suit our requirements. For example, the publicly available \textit{Events-2012 Twitter dataset}~\cite{mcminn2013building} contains generic event descriptions such as \textit{Politics}, \textit{Sports}, \textit{Culture} etc. Our work targets more specific event categories~\cite{bhardwaj2019TKDE}.
Following previous studies \cite{ritter2015weakly}, we collect event-related microposts from Twitter using 11 and 8 seed events~(see Section \ref{sec:model}) for \emph{CyberAttack} and \emph{PoliticianDeath}, respectively. Unlabeled microposts are collected by using the keyword \lq hack' for \emph{CyberAttack}, while for \emph{PoliticianDeath}, we use a set of keywords related to \lq politician' and \lq death' (such as \lq bureaucrat', \lq dead' etc.) 
For each dataset, we randomly select 500 tweets from the unlabeled subset and manually label them for evaluation. Table~\ref{tab:stats} shows key statistics from our two datasets.

\begin{table}[!t]
\centering
\small
\caption{Statistics of the datasets in our experiments.}\label{tab:stats}
\begin{tabular}{lccc}
\toprule
\textbf{Dataset} & \textbf{\#Positive} & \textbf{\#Unlabeled} & \textbf{\#Test} \\
\midrule
CyberAttack  & 2,600 & 86,000 & 500 \\
PoliticianDeath & 900  & 7,000 & 500\\
\bottomrule
\end{tabular}
\end{table}

\smallskip
\noindent\textbf{Comparison Methods.} To demonstrate the generality of our approach on different event detection models, we consider Logistic Regression (LR)~\cite{ritter2015weakly} and Multilayer Perceptron (MLP)~\cite{chambers2018detecting} 
as the target models. As the goal of our experiments is to demonstrate the effectiveness of our approach as a new model training technique, we use these widely used models. Also, we note that in our case other neural network models with more complex network architectures for event detection, such as the bi-directional LSTM~\cite{chang2016expectation}, turn out to be less effective than a simple feedforward network. For both LR and MLP, we evaluate our proposed human-AI loop approach for keyword discovery and expectation estimation by comparing against the weakly supervised learning method proposed by~\citeauthor{ritter2015weakly}~(\citeyear{ritter2015weakly}) and~\citeauthor{chang2016expectation}~(\citeyear{chang2016expectation}) where only one initial keyword is used with an expectation estimated by an individual expert.

\smallskip
\noindent\textbf{Parameter Settings.} We empirically set optimal parameters based on a held-out validation set that contains 20\% of the test data. These include the hyperparamters of the target model, those of our proposed probabilistic model, and the parameters used for training the target model. We explore MLP with 1, 2 and 3 hidden layers and apply a grid search in {32, 64, 128, 256, 512} for the dimension of the embeddings and that of the hidden layers. 
For the coefficient of expectation regularization, we follow \citeauthor{mann2007simple} (\citeyear{mann2007simple}) and set it to $\lambda=10 \times$ \#labeled examples. 
For model training, we use the Adam \cite{kingma2014adam} optimization algorithm for both models.

\begin{table*}[btp]
\centering
\caption{Performance of the target models trained by our proposed human-AI loop approach on the experimental datasets at different iterations. Results are given in percentage. }
\label{tab:res}
\begin{tabular}{lcccccccccccc}
\toprule
\multicolumn{1}{l}{\multirow{2}{*}{\textbf{Dataset}}} & \multicolumn{1}{l}{\multirow{2}{*}{\textbf{Method}}} & \multirow{2}{*}{\textbf{Metric}} & \multicolumn{9}{c}{\textbf{Iteration}}                                         \\ \cline{4-12} 
\multicolumn{1}{c}{} & \multicolumn{1}{c}{} & & 1     & 2     & 3     & 4     & 5     & 6     & 7     & 8     & 9     \\ \midrule
\multirow{4}{*}{Cyber Attack}                & \multirow{2}{*}{LR}                         
& AUC       & 66.69 & 72.67 & 69.02 & 69.18 & 70.41 & 70.22 & 70.66 & 70.66 &  70.53 \\ 
& 
& Accuracy  & 71.04 & 74.07 & 74.07 & 74.07 & 72.72 & 72.72 & 72.72 & 72.72 & 72.39  \\ \cline{2-12} 
                                             & \multirow{2}{*}{MLP}                        
& AUC       & 60.79 & 66.06 & 70.5 & 72.83 & 76.06 & 75.28 & 75.98 & 75.60 & 75.81      \\ 
                                             &                                             
& Accuracy  & 70.37 & 73.06 & 73.06 & 73.40  & 75.42 & 75.08 & 75.42 & 74.41 & 75.75  \\ \hline
\multirow{4}{*}{Politician Death}           & \multirow{2}{*}{LR} 
& AUC    & 49.37 & 60.69 & 61.32 & 63.45 & 62.71 & 62.72 & 63.07 & 63.50  & 64.68 \\ 
&  
& Accuracy  & 76.53 & 82.65 & 83.67 & 83.67 & 82.99 & 83.33 & 82.99 & 82.99 & 82.99  \\ \cline{2-12} 

                                             & \multirow{2}{*}{MLP}                        
& AUC       & 56.81 & 74.20 & 72.60 & 73.80 & 72.59 & 73.00 & 76.11  & 76.52 & 77.17   \\ 
                                             &                                             
& Accuracy  & 76.53 & 87.07 & 86.05 & 87.07  & 85.71 & 86.05 & 86.39 & 87.07 & 87.07    \\ \bottomrule
\end{tabular}
\end{table*}

\smallskip
\noindent\textbf{Evaluation.}
Following~\citeauthor{ritter2015weakly}~(\citeyear{ritter2015weakly}) and~\citeauthor{konovalov2017learning}~(\citeyear{konovalov2017learning}), we use accuracy and area under the precision-recall curve (AUC) metrics to measure the performance of our proposed approach. 
We note that due to the imbalance in our datasets (20\% positive microposts in \emph{CyberAttack} and 27\% in \emph{PoliticianDeath}), accuracy is dominated by negative examples; AUC, in comparison, better characterizes the discriminative power of the model.

\smallskip
\noindent\textbf{Crowdsourcing.} We chose Level 3 workers on the Figure-Eight\footnote{https://www.figure-eight.com/} crowdsourcing platform for our experiments. The inter-annotator agreement in micropost classification is taken into account through the EM algorithm. For keyword discovery, we filter keywords based on the frequency of the keyword being selected by the crowd. In terms of cost-effectiveness, our approach is motivated from the fact that crowdsourced data annotation can be expensive, and is thus designed with minimal crowd involvement. For each iteration, we selected ~50 tweets for keyword discovery and ~50 tweets for micropost classification per keyword. For a dataset with 80k tweets (e.g., \emph{CyberAttack}), our approach only requires to manually inspect 800 tweets (for 8 keywords), which is only 1\% of the entire dataset.

\subsection{Results of our Human-AI Loop (Q1)}
Table~\ref{tab:res} reports the evaluation of our approach on both the \emph{CyberAttack} and \emph{PoliticianDeath} event categories. Our approach is configured such that each iteration starts with 1 new keyword discovered in the previous iteration. 


Our approach improves LR by 5.17\% (Accuracy) and 18.38\% (AUC), and MLP by 10.71\% (Accuracy) and 30.27\% (AUC) on average. Such significant improvements clearly demonstrate that our approach is effective at improving model performance. We observe that the target models generally converge between the 7\textsuperscript{th} and 9\textsuperscript{th} iteration on both datasets when performance is measured by AUC. 
The performance can slightly degrade when the models are further trained for more iterations on both datasets. This is likely due to the fact that over time, the newly discovered keywords entail lower novel information for model training. For instance, for the \emph{CyberAttack} dataset the new keyword in the 9\textsuperscript{th} iteration \lq \textit{election}' frequently co-occurs with the keyword \lq \textit{russia}' in the 5\textsuperscript{th} iteration (in microposts that connect Russian hackers with US elections), thus bringing limited new information for improving the model performance. As a side remark, we note that the models converge faster when performance is measured by accuracy. Such a comparison result confirms the difference between the metrics and shows the necessity for more keywords to discriminate event-related microposts from non event-related ones.

\subsection{Comparative Results on Keyword Discovery (Q2)}

\begin{figure}[t!]
        \centering
        \subfigure[CyberAttack]{
            \includegraphics[width=0.34\textwidth]{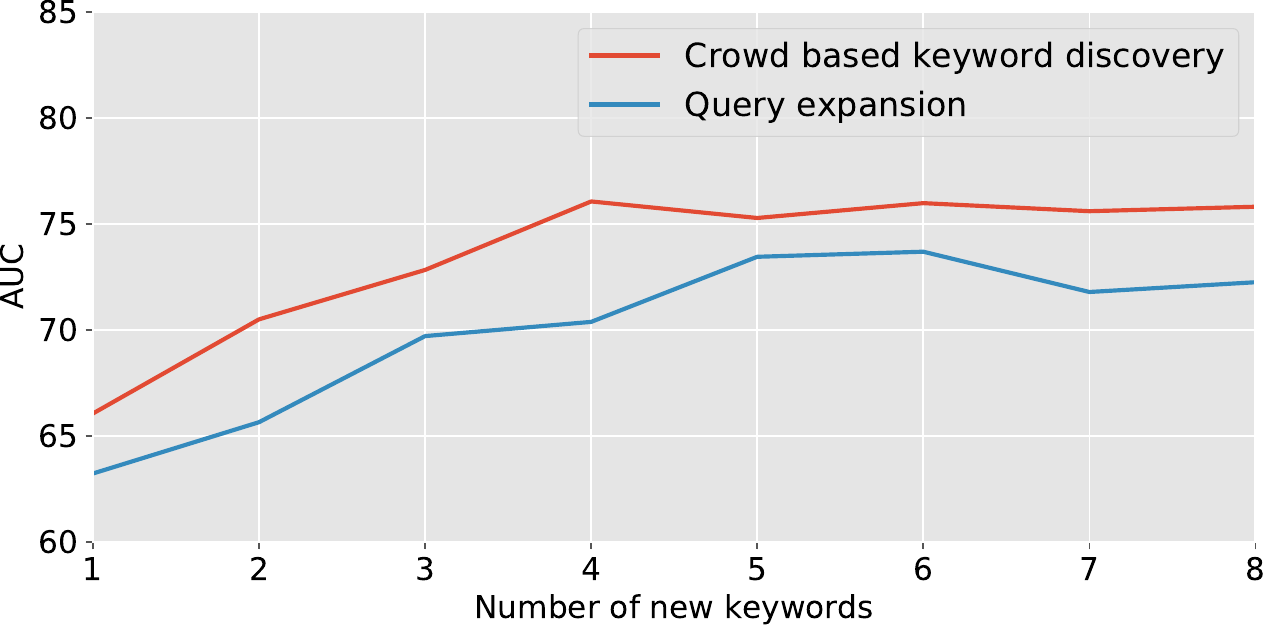}
        }
        \subfigure[PoliticianDeath]{
            \includegraphics[width=0.34\textwidth]{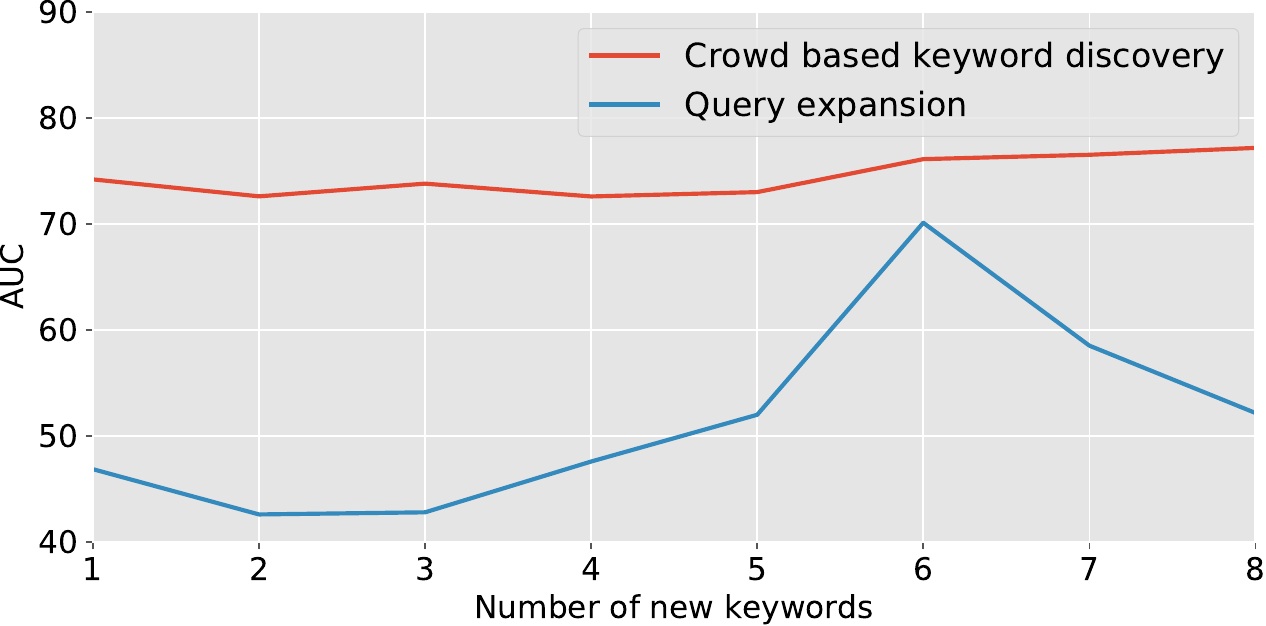}
        }
        \caption{Comparison between our keyword discovery method and query expansion method for MLP (similar results for LR).}
        \label{fig:Q2}
\end{figure}

Figure~\ref{fig:Q2} shows the evaluation of our approach when discovering new informative keywords for model training~(see Section \ref{sec:model}:~\emph{Keyword Discovery}). We compare our human-AI collaborative way of discovering new keywords against a query expansion~(QE) approach~\cite{diaz2016query,kuzi2016query} that leverages word embeddings to find similar words in the latent semantic space. Specifically, we use pre-trained word embeddings based on a large Google News dataset\footnote{https://code.google.com/archive/p/word2vec/} for query expansion. For instance, the top keywords resulting from QE for \lq politician' are, \lq deputy',\lq ministry',\lq secretary', and \lq minister'. For each of these keywords, we use the crowd to label a set of tweets and obtain a corresponding expectation.

We observe that our approach consistently outperforms QE by an average of $4.62\%$ and $52.58\%$ AUC on \emph{CyberAttack} and \emph{PoliticianDeath}, respectively. The large gap between the performance improvements for the two datasets is mainly due to the fact that microposts that are relevant for \emph{PoliticianDeath} are semantically more complex than those for \emph{CyberAttack}, as they encode noun-verb relationship (e.g., ``the \emph{king} of ... \emph{died} ...'') rather than a simple verb (e.g., ``... \emph{hacked}.'') for the \emph{CyberAttack} microposts. QE only finds synonyms of existing keywords related to either \lq politician' \emph{or} \lq death', however cannot find a meaningful keyword that fully characterizes the death of a politician. For instance, QE finds the keywords \lq kill' and \lq murder', which are semantically close to \lq death' but are not specifically relevant to the death of a politician. Unlike QE, our approach identifies keywords that go beyond mere synonyms and that are more directly related to the end task, i.e., discriminating event-related microposts from non related ones. Examples are \lq demise' and \lq condolence'. As a remark, we note that in Figure~\ref{fig:Q2}(b), the increase in QE performance on \emph{PoliticianDeath} is due to the keywords \lq deputy' and \lq minister', which happen to be highly indicative of the death of a politician in our dataset; these keywords are also identified by our approach.

\subsection{Cost-Effectiveness Results (Q3)}
To demonstrate the cost-effectiveness of using crowdsourcing for obtaining new keywords and consequently, their expectations, we compare the performance of our approach with an approach using crowdsourcing to only label microposts for model training at the same cost.
Specifically, we conducted an additional crowdsourcing experiment where the same cost used for keyword discovery in our approach is used to label additional microposts for model training.
These newly labeled microposts are used with the microposts labeled in the micropost classification task of our approach~(see Section \ref{sec:model}:~\emph{Micropost Classification}) and the expectation of the initial keyword to train the model for comparison. 
The model trained in this way increases AUC by 0.87\% for \emph{CyberAttack}, and by 1.06\% for \emph{PoliticianDeath}; in comparison, our proposed approach increases AUC by 33.42\% for \emph{PoliticianDeath} and by 15.23\% for \emph{CyberAttack} over the baseline presented by~\citeauthor{ritter2015weakly}).
These results show that using crowdsourcing for keyword discovery is significantly more cost-effective than simply using crowdsourcing to get additional labels when training the model.


\subsection{Expectation Inference Results (Q4)}
\begin{figure}
       \centering
        \subfigure[CyberAttack]{
            \includegraphics[width=0.34\textwidth]{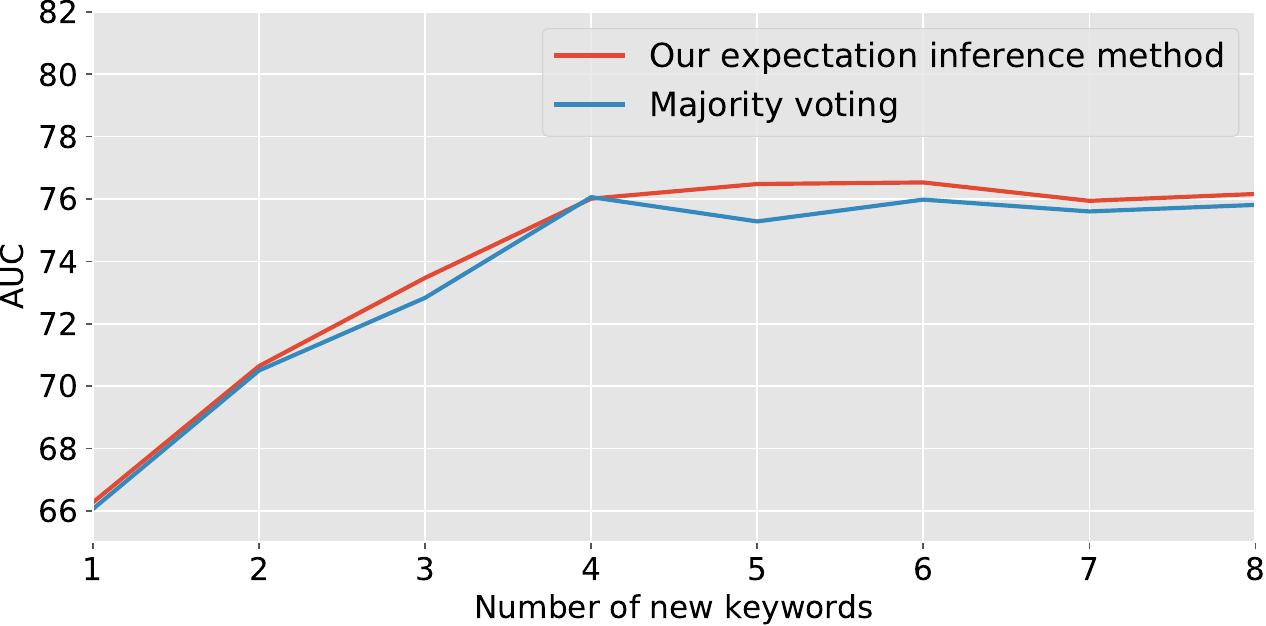}
        }
        \subfigure[PoliticianDeath]{
            \includegraphics[width=0.34\textwidth]{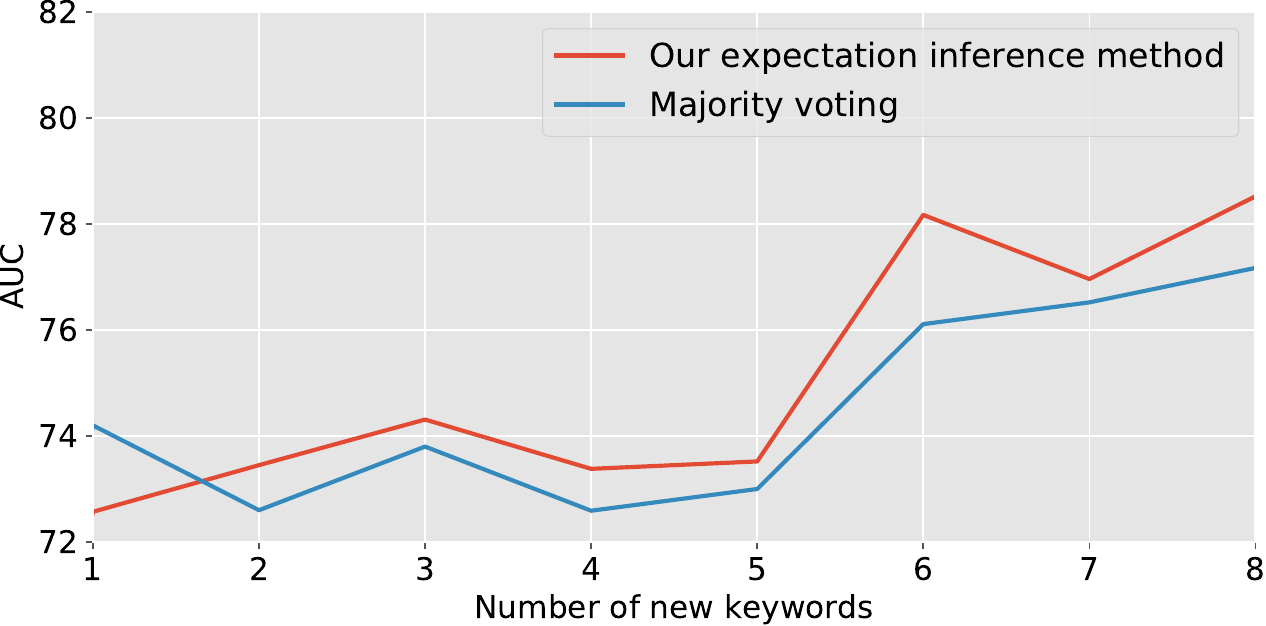}
        }
        \caption{Comparison between our expectation inference method and majority voting for MLP (similar results for LR).}
        \label{fig:Q3}
\end{figure}

To investigate the effectiveness of our expectation inference method, we compare it against a majority voting approach, a strong baseline in truth inference \cite{zheng2017truth}. Figure~\ref{fig:Q3} shows the result of this evaluation. We observe that our approach results in better models for both \emph{CyberAttack} and \emph{PoliticianDeath}. Our manual investigation reveals that workers' annotations are of high reliability, which explains the relatively good performance of majority voting. Despite limited margin for improvement, our method of expectation inference improves the performance of majority voting by $0.4\%$ and $1.19\%$ AUC on \emph{CyberAttack} and \emph{PoliticianDeath}, respectively.

\section{Related Work}
\label{sec:related}

\noindent\textbf{Event Detection.} 
The techniques for event extraction from microblogging platforms can be classified according to their domain specificity and their detection method~\cite{atefeh2015survey}. Early works mainly focus on open domain event detection~\cite{benson2011event,ritter2012open,chierichetti2014event}. Our work falls into the category of domain-specific event detection~\cite{bhardwaj2019TKDE}, which has drawn increasing attention due to its relevance for various applications such as cyber security~\cite{ritter2015weakly,chambers2018detecting} and public health~\cite{akbari2016tweets,lee2017adverse}. In terms of  technique, our proposed detection method is related to the recently proposed weakly supervised learning methods~\cite{ritter2015weakly,chang2016expectation,konovalov2017learning}. This comes in contrast with fully-supervised learning methods, which are often limited by the size of the training data (e.g., a few hundred examples)~\cite{sakaki2010earthquake,sadri2016online}.

\smallskip
\noindent\textbf{Human-in-the-Loop Approaches.} Our work extends weakly supervised learning methods by involving humans in the loop~\cite{vaughan2017making}. Existing human-in-the-loop approaches mainly leverage crowds to label individual data instances~\cite{yan2011active,yang2018leveraging} or to debug the training data~\cite{krishnan2016activeclean,yang2019scalpel} or components~\cite{parikh2011human,mottaghi2013analyzing,nushi2017human} of a machine learning system. Unlike these works, we leverage crowd workers to label sampled microposts in order to obtain keyword-specific expectations, which can then be generalized to help classify microposts containing the same keyword, thus amplifying the utility of the crowd. Our work is further connected to the topic of interpretability and transparency of machine learning models~\cite{ribeiro2016should,lipton2016mythos,doshi2017towards}, for which humans are increasingly involved, for instance for post-hoc evaluations of the model's interpretability. In contrast, our approach directly solicits informative keywords from the crowd for model training, thereby providing human-understandable explanations for the improved model.

\section{Conclusion}
\label{sec:conclusion}

In this paper, we presented a new human-AI loop approach for keyword discovery and expectation estimation to better train event detection models. Our approach takes advantage of the disagreement between the crowd and the model to discover informative keywords and leverages the joint power of the crowd and the model in expectation inference. We evaluated our approach on real-world datasets and showed that it significantly outperforms the state of the art and that it is particularly useful for detecting events where relevant microposts are semantically complex, e.g., the death of a politician. As future work, we plan to parallelize the crowdsourcing tasks and optimize our pipeline in order to use our event detection approach in real-time.

\section{Acknowledgements}
This project has received funding from the Swiss National Science Foundation (grant \#407540\_167320 Tighten-it-All) and from the European Research Council (ERC) under the European Union's Horizon 2020 research and innovation programme (grant agreement 683253/GraphInt).

\begin{quote}
\bibliographystyle{aaai}
\bibliography{aaai.bib}
\end{quote}
\end{document}